\documentclass[superscriptaddress,aps,twocolumn,prl,tightenlines,floatfix,showpacs,amsmath,amssymb]{revtex4-2}

\usepackage{dsfont}
\usepackage{graphicx}
\usepackage{pdfpages}
\makeatletter
\AtBeginDocument{\let\LS@rot\@undefined}
\makeatother

\begin{document}

\title{Topological Phase Transitions of Dirac Magnons in Honeycomb Ferromagnets}

\author{Yu-Shan Lu}
\author{Jian-Lin Li}
\affiliation{Department of Electrophysics, National Yang Ming Chiao Tung University, Hsinchu, Taiwan}
\author{Chien-Te Wu}
\affiliation{Department of Electrophysics, National Yang Ming Chiao Tung University, Hsinchu, Taiwan}
\affiliation{Physics Division, National Center for Theoretical Sciences, Taipei, Taiwan}

\begin{abstract}
The study of the magnonic thermal Hall effect in magnets with Dzyaloshinskii-Moriya interaction  (DMI) has recently drawn attention because of the underlying topology.
Topological phase transitions may arise when there exist two or more distinct topological phases, and they are often revealed by a gap-closing phenomenon. In this work, we consider the magnons in honeycomb ferromagnets described by a Heisenberg Hamiltonian containing both an out-of-plane DMI and a Zeeman interaction. 
We demonstrate that the magnonic system exhibits temperature (or magnetic field) driven topological phase transitions due to magnon-magnon interactions.
Specifically, when the temperature increases, the magnonic energy gap at Dirac points closes and reopens at a critical temperature, $T_c$. By showing that the Chern numbers of the magnonic bands are distinct above and below $T_c$, we confirm that the gap-closing phenomenon is indeed a signature for the topological phase transitions. Furthermore,  our analysis indicates that the thermal Hall conductivity in the magnonic system exhibits a sign reversal at $T_c$, which can serve as an experimental probe of its topological nature. Our theory predicts that in $\rm{CrI_3}$ such a phenomenon exists and is experimentally accessible.
\end{abstract}

\maketitle

\textit{Introduction.$-$} 
In condensed matter physics, the notion of topologically-ordered phases has opened new avenues to understand exotic phases of matter
that cannot fit into the Landau's paradigm. The integer quantum Hall effect in a two-dimensional electron gas is one such example in which the time-reversal symmetry is broken and the  system is in a topologically-ordered phase characterized by a quantized Hall conductance. Another prominent example is the discovery of symmetry-protected topological orders in topological insulators~\cite{Kane2005,Bernevig2006,Kane2005b}.  While these ordered phases are not classified in the Landau's theory, they can be linked to certain topological invariants. Apart from the above examples concerning fermions, the topological aspect of bosons has also received considerable interest, e.g., a symmetry-protected topological phase of ${}^{87}\rm{Rb}$ atoms trapped in a one-dimensional lattice has been realized experimentally~\cite{leseleuc}.
Also, the study of topological orders in systems consisting of bosonic collective excitations such as magnons is another important subject because it connects with the possibility of a nonzero thermal Hall conductivity~\cite{PhysRevB.97.180401,PhysRevB.97.134411,PhysRevB.97.054404,PhysRevB.97.174407,PhysRevB.97.174413,Owerre_2018,PhysRevB.101.195133,PhysRevB.103.014407,Owerre2018,Banerjee_2020,PhysRevB.91.125413}. 
Unlike electrons, magnons are charge neutral, and thereby not experiencing any Lorentz force. Therefore, the topology of magnonic systems cannot be studied via a standard Hall effect. However,
it is shown~\cite{PhysRevLett.106.197202} that in magnonic systems an antisymmetric interaction such as  Dzyaloshinskii-Moriya interaction (DMI), between spin moments plays a similar role as the Lorentz force  and gives rise to a thermal Hall effect.

\par 
The theoretical work on the thermal Hall effect in quantum magnets with particular lattice structures such as kagome lattice was 
first studied in Ref.~\cite{PhysRevLett.104.066403}. 
Experimental evidence for the magnonic thermal Hall effect in a ferromagnetic insulator with pyrochlore lattice structure was also reported around the same time~\cite{Onose297}. In addition to pyrochlore, 
experiments on the thermal Hall effect of Cu(1-3,bdc) with kagome lattice also reveal a magnon origin~\cite{PhysRevLett.115.106603}. 
Interestingly, the observed thermal Hall conductivity in Ref.~\cite{PhysRevLett.115.106603} undergoes an unusual sign reversal by tuning the temperature.
The origin of the sign reversal 
can be understood by examining the topological properties of the magnons in kagome ferromagnets~\cite{PhysRevB.91.125413,PhysRevB.89.134409,PhysRevB.90.024412}. 
In essence, the  contributions from the lowest and higher magnonic bands to transport depend on the thermal population and Berry curvature of these bands. Because the Berry curvature of these bands may have opposite signs, the sign of the thermal Hall conductivity can change with the temperature without a topological phase transition. 




\par 
The magnonic thermal Hall effect is also theoretically predicted to exist in honeycomb magnets~\cite{Owerre_2016,doi:10.1063/1.4959815} in addition to widely studied kagome magnets~\cite{PhysRevB.89.134409,PhysRevB.103.054405,PhysRevB.97.134411,PhysRevLett.115.106603}.
It is found that Dirac points are gapped as the orbital time-reversal symmetry is broken by a DMI, similar to the effect of the next-nearest-neighbor interaction in the Haldane model~\cite{PhysRevLett.61.2015,PhysRevLett.117.227201}. 
In contrast to kagome lattices, the sign of the thermal Hall conductivity in magnets with honeycomb lattice is predicted~\cite{doi:10.1063/1.4959815} to never change for all values of relevant parameters considered such as the temperature and the strength of an applied magnetic field. There, many-body interaction effects are neglected. 
Recently, Pershoguba {\it et al.}~\cite{PhysRevX.8.011010} considered magnon-magnon interactions in honeycomb ferromagnets and found that these many-body interactions lead to a notable momentum-dependent renormalization of  the magnon bands. Their theory successfully resolved anomalies in the neutron-scattering data for $\text{CrBr}_{3}$ that had been unexplained for nearly fifty years. It is to be noted that  DMI is not taken into account in 
their theory, although its presence in $\rm{CrBr_3}$ and $\rm{CrI_3}$ has recently been experimentally confirmed~\cite{PhysRevB.104.L020402,PhysRevX.8.041028}.

The direction of DMI is essential in determining both the topological properties and types of magnon-magnon interaction in honeycomb ferromagnets~\cite{PhysRevX.11.021061}. 
The nearest-neighbor (NN) DMI is absent and the next-nearest-neighbor (NNN) DMI vector is allowed and along the out-of-plane direction according to the Moriya's rules~\cite{moriya}. This is the situation we consider here. When
structural inversion asymmetry is present, an in-plane DMI between nearest neighbors is permitted  and generates particle-number-nonconserving processes. This in-plane DMI, along with an out-of-plane magnetic field, leads to important topological phenomena such as 
the existence of chiral edge states~\cite{PhysRevX.11.021061}. 

From the above discussion, it is clear that both DMI and many-body effects play an important role and cannot be ignored. Therefore, in this paper, we incorporate  DMI and magnon-magnon interactions into our model Hamiltonian to investigate the topological aspect of honeycomb ferromagnets. We find that topological phase transitions in  honeycomb ferromagnets 
are driven by the magnon-magnon interactions and these transitions are marked by the sign change of the thermal Hall conductivity at a finite magnetic field. It is in sharp contrast to the previous theoretical results on kagome ferromagnets where the sign flip is not accompanied with a topological phase transition~\cite{PhysRevB.91.125413,PhysRevB.89.134409,PhysRevB.90.024412}. Below we shall first introduce our theoretical formalism and then present our results on the topological properties of honeycomb ferromagnets.

\begin{figure}[h] 
\includegraphics[width=3.0in,clip]{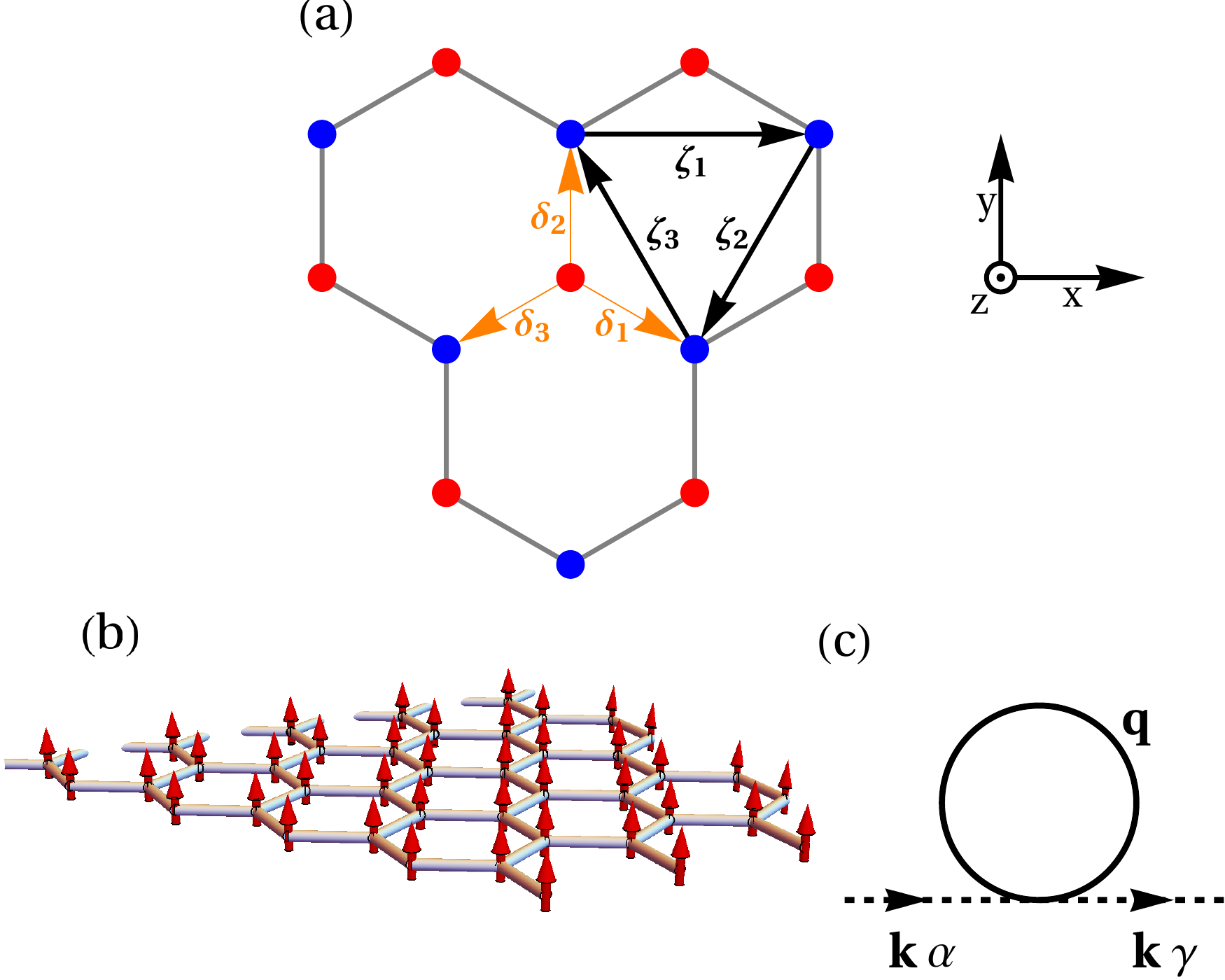}
\caption{
(a) Schematics of a honeycomb lattice. Three nearest-neighbor and next-nearest-neighbor vectors are labeled as $\boldsymbol{\delta}_n$ and $\boldsymbol{\zeta}_n$ ($n=1,2,3$), respectively. (b) The localized spins are represented by red arrows and point along the $z$-axis, same as the external magnetic field, in the $T=0$ limit. (c) The Feynman diagram for the Hartree-type self-energy. Here the Green's functions to be contracted are denoted by the dashed arrows with $\alpha$ and $\gamma$ representing the up or down band.
}
\label{fig:1}
\end{figure}

\textit{Background theory.$-$}We consider localized spin moments arranged on a honeycomb lattice in the $xy$ plane as shown in Fig.~\ref{fig:1}(a).
The corresponding Hamiltonian associated with these spins is
\begin{equation}
    \label{Hamiltonian}
    \begin{split}
    H=&-J\sum_{\langle ij \rangle}\hat{\mathbf{S}}_{i}\cdot\hat{\mathbf{S}}_{j}
    -J^\prime\sum_{\langle\langle ij \rangle\rangle}\hat{\mathbf{S}}_{i}\cdot\hat{\mathbf{S}}_{j}\\
    &+\sum_{\langle\langle ij \rangle\rangle}\mathbf{D}_{ij}\cdot\hat{\mathbf{S}}_{i}\times\hat{\mathbf{S}}_{j}-g\mu_BB\hat{\mathbf{z}}\cdot\sum_{i}\hat{\mathbf{S}}_{i},
    \end{split}
\end{equation}
where the first two terms are the NN and NNN ferromagnetic Heisenberg exchange coupling, i.e., $J,J^\prime>0$.
The third term is an out-of-plane NNN Dzyaloshinskii-Moriya interaction with $\mathbf{D}_{ij}=\nu_{ij}D\hat{\mathbf{z}}$, where   $\nu_{ij}=+1(-1)$ for  clockwise(counterclockwise) hopping.
The last term is a Zeeman term. We define $h\equiv g\mu_B B$, where $g$ is the $g$-factor, $\mu_B$ is the Bohr magneton, and $B\hat{\mathbf{z}}$ is an out-of-plane magnetic field. Because of the presence of the Zeeman field, the magnons are no longer protected by the Goldstone theorem and the spectrum is gapped at the $\Gamma$ point~\cite{Owerre_2016}. 
The magnetic field also tends to align the localized spins along its direction as depicted in Fig.~\ref{fig:1}(b).
\par We convert the spin operators in Eq.~(\ref{Hamiltonian}) into magnon creation ($\hat{c}^\dagger$) and annihilation ($\hat{c}$) operators by using the Holstein-Primakoff (HP) transformations: $\hat{S}_{x}+i\hat{S}_{y} = \sqrt{2S-\hat{c}^{\dagger} \hat{c}}\hat{c}$, $\hat{S}_{x}-i\hat{S}_{y} =\hat{c}^{\dagger} \sqrt{2S-\hat{c}^{\dagger} \hat{c}}$, and $\hat{S}_{z}=S-\hat{c}^{\dagger}\hat{c}$. In the low temperature limit, $2S\gg\langle\hat{n}\rangle=\langle\hat{c}^\dagger\hat{c}\rangle$, and the square roots can be expanded in powers of $1/\sqrt{S}$. Truncated to $S^{\frac{1}{2}}$, we obtain the noninteracting Hamiltonian in the momentum space, $\mathcal{H}_{0}=\sum_{\mathbf{k}}\boldsymbol{\Psi^{\dagger}_{\mathbf{k}}}H_0(\mathbf{k})\boldsymbol{\Psi_{\mathbf{k}}},$
where $\boldsymbol{\Psi^{\dagger}_{\mathbf{k}}}=(\hat{a}^{\dagger}_{\mathbf{k}}\ \hat{b}^{\dagger}_{\mathbf{k}})$ is a spinor denoting the degrees of freedom for the two sublattices 
and $H_0(\mathbf{k})=h_{0}(\mathbf{k})\sigma_{0}+h_{x}(\mathbf{k})\sigma_{x}-h_{y}(\mathbf{k})\sigma_{y}+h_{z}(\mathbf{k})\sigma_{z}$. 
Here, $h_{0}(\mathbf{k})=v_{0}-2v_{t}\cos{\phi} p_{\mathbf{k}}$, $h_{x}(\mathbf{k})=-v_{s}\rm{Re}\left(\gamma_{\bf{k}}\right)$, $h_{y}(\mathbf{k})=-v_{s}\rm{Im}\left(\gamma_{\bf{k}}\right)$, and $h_{z}(\mathbf{k})=2v_{t}\sin{\phi}\rho_{\mathbf{k}}$. In the above expressions, $\gamma_{\bf k}\equiv\sum_{n=1}^3e^{i\bf{k}\cdot{\boldsymbol{\delta}}_n}$ with $\boldsymbol{\delta}_n$ being the three NN vectors, $v_{0}=3v_{s}+6v_{s}^\prime+h$, $v_{t}=\sqrt{v_{s}^{\prime2}+v_{D}^{2}}$, $v_{s}(v_{s^{'}})(v_{D})=JS(J^{'}S)(DS)$, $\phi=\text{arctan}(D/J')$, $p_{\mathbf{k}}=\sum_{n=1}^3\cos{(\mathbf{k}\cdot\boldsymbol{\zeta}_{n})}$, and $\rho_{\mathbf{k}}=\sum_{n=1}^3\sin{(\mathbf{k}\cdot\boldsymbol{\zeta}_{n})}$, where $\boldsymbol{\zeta}_{n}$ are the three NNN vectors.
Diagonalizing $H_0\left(\bf{k}\right)$,
we obtain the single-particle magnon energies,
\begin{equation}
    \label{single-particle_magnon_energies}
    \varepsilon_{\alpha}(\mathbf{k}) = h_{0}(\mathbf{k}) + \lambda \epsilon(\mathbf{k}),
\end{equation}
where $\epsilon(\mathbf{k})=\sqrt{h_{x}^2(\mathbf{k})+h_{y}^2(\mathbf{k})+h_{z}^2(\mathbf{k})}$ and $\lambda=1(-1)$ for the up(down) band, $\alpha=u(\alpha=d)$.
The corresponding wavefunction is given by
\begin{equation}
    \label{Blochwave_of_magnon}
    |\psi_{\alpha}(\mathbf{k})\rangle=\frac{1}{\sqrt{2}} 
    \begin{pmatrix}
    &\sqrt{1+\lambda\frac{h_{z}(\mathbf{k})}{\epsilon(\mathbf{k})}} \\
    &\lambda e^{-i\Phi(\mathbf{k})}\sqrt{1-\lambda\frac{h_{z}(\mathbf{k})}{\epsilon(\mathbf{k})}}
    \end{pmatrix},
\end{equation}
where $\Phi({\mathbf{k}})=\text{arg}\left[h_{x}(\mathbf{k})+ih_{y}(\mathbf{k})\right]$. From Eq.~(\ref{single-particle_magnon_energies}), it is easy to see that the spectrum remains gapped
at the Dirac points $\mathbf{K}_{\pm}$ with a gap size of $2|h_{z}\left(\mathbf{K}_\pm\right)|$ provided that the DMI is nonzero.
\par The Berry curvature
of each energy band can be computed by $\boldsymbol{\Omega}_{\alpha}(\mathbf{k})=\nabla_{\mathbf{k}}\times\langle\psi_{\alpha}(\mathbf{k})|i\nabla_{\mathbf{k}}|\psi_{\alpha}(\mathbf{k})\rangle$.
The topological phases of our system are characterized by the Chern numbers of each band, defined as the integration of the corresponding Berry curvature over the Brillouin zone (BZ), $C_{\alpha} = \frac{1}{2\pi} \int_{\rm BZ} d^{2}k \Omega_{z,\alpha}(\mathbf{k})$.
For the up and down bands of the noninteracting Hamiltonian, 
the Chern numbers 
are -1 and +1, respectively.
The thermal Hall conductivity is also related to the Berry curvatures by the following expression derived in Ref.~\cite{PhysRevLett.106.197202},
\begin{equation}
    \label{Hallconductivity}
    \kappa_{xy}(T)=-\frac{k_{B}^{2}T}{(2\pi)^{2}\hslash} \sum\limits_{\alpha}\int_{BZ}d^{2}k c_{2}(n_{\alpha})\Omega_{z, \alpha}(\mathbf{k}),
\end{equation}
where $n_{\alpha}=[ e^{\beta\varepsilon_{\alpha}(\mathbf{k})}-1]^{-1}$ is the Bose distribution function and $\beta=1/{k_B T}$, $c_{2}(x)=(1+x)(\log\frac{1+x}{x})^{2}-(\log x)^{2}-2\text{Li}_{2}(-x)$, and $\text{Li}_{2}$ is the dilogarithm. 
We will demonstrate later that the sign of the thermal Hall conductivity is the same as that of the Chern number for the up band. 
For the noninteracting Hamiltonian, the system is always in the same topological phase  $(C_u=-1,\,C_d=+1)$ and $\kappa_{xy}(T)$ stays negative~\cite{doi:10.1063/1.4959815}  for all parameter regimes as long as the magnetic field  is not flipped.
\par Next, we consider the next order  in $S^{-\frac{1}{2}}$ of the HP transformation to obtain magnon-magnon interactions. Using the eigenstates of the noninteracting Hamiltonian, we express the interacting part $\cal{H}_{\rm int}$ of the full Hamiltonian as,
\begin{equation}
\label{H_4}
    \begin{split}
    \cal{H}_{\rm int}=&\frac{1}{4SN}\sum\limits_{\{\mathbf{k}_i\}}\left[V_{1,\{\mathbf{k}_i\}}\hat{u}^{\dagger}_{\mathbf{k}_{1}}\hat{u}^{\dagger}_{\mathbf{k}_{2}}\hat{u}_{\mathbf{k}_{3}}\hat{u}_{\mathbf{k}_{4}}\right.\\
     +&V_{2,\{\mathbf{k}_i\}}\hat{u}^{\dagger}_{\mathbf{k}_{1}}\hat{u}^{\dagger}_{\mathbf{k}_{2}}\hat{u}_{\mathbf{k}_{3}} \hat{d}_{\mathbf{k}_{4}}+V_{3,\{\mathbf{k}_i\}}\hat{u}^{\dagger}_{\mathbf{k}_{1}}\hat{u}^{\dagger}_{\mathbf{k}_{2}}\hat{d}_{\mathbf{k}_{3}}\hat{d}_{\mathbf{k}_{4}}\\
    +&V_{4,\{\mathbf{k}_i\}}\hat{u}^{\dagger}_{\mathbf{k}_{1}}\hat{d}^{\dagger}_{\mathbf{k}_{2}}\hat{u}_{\mathbf{k}_{3}} \hat{u}_{\mathbf{k}_{4}}+V_{5,\{\mathbf{k}_i\}}\hat{u}^{\dagger}_{\mathbf{k}_{1}}\hat{d}^{\dagger}_{\mathbf{k}_{2}}\hat{u}_{\mathbf{k}_{3}}\hat{d}_{\mathbf{k}_{4}}\\
    +&V_{6,\{\mathbf{k}_i\}}\hat{u}^{\dagger}_{\mathbf{k}_{1}}\hat{d}^{\dagger}_{\mathbf{k}_{2}}\hat{d}_{\mathbf{k}_{3}} \hat{d}_{\mathbf{k}_{4}}+V_{7,\{\mathbf{k}_i\}}\hat{d}^{\dagger}_{\mathbf{k}_{1}}\hat{d}^{\dagger}_{\mathbf{k}_{2}}\hat{u}_{\mathbf{k}_{3}}\hat{u}_{\mathbf{k}_{4}}\\
    +&\left.V_{8,\{\mathbf{k}_i\}}\hat{d}^{\dagger}_{\mathbf{k}_{1}}\hat{d}^{\dagger}_{\mathbf{k}_{2}} \hat{u}_{\mathbf{k}_{3}} \hat{d}_{\mathbf{k}_{4}}+V_{9,\{\mathbf{k}_i\}}\hat{d}^{\dagger}_{\mathbf{k}_{1}}\hat{d}^{\dagger}_{\mathbf{k}_{2}}\hat{d}_{\mathbf{k}_{3}}\hat{d}_{\mathbf{k}_{4}}\right]\\
    &\delta_{-\mathbf{k}_{1}-\mathbf{k}_{2}+\mathbf{k}_{3}+\mathbf{k}_{4},0},
    \end{split}
\end{equation}
where ${\{\mathbf{k}_i\}}=\{\mathbf{k}_{1},\mathbf{k}_{2},\mathbf{k}_{3},\mathbf{k}_{4}\}$ and $N$ is the total number of sublattice sites. $\cal{H}_{\rm{int}}$ is expressed in the $ud$ basis related to the $ab$ basis by the relation $(\hat{u}_{\mathbf{k}}\;\hat{d}_{\mathbf{k}})^T=P^\dagger(\hat{a}_{\mathbf{k}}\;\hat{b}_{\mathbf{k}})^T$, where $P$ is the unitary transformation matrix associated with Eq.~(\ref{Blochwave_of_magnon}). 
The process of obtaining $V_{1,\{\mathbf{k}_i\}}$ to $V_{9,\{\mathbf{k}_i\}}$ is given in the supplementary material. To properly take into account many-body correlation effects from the magnon-magnon interactions, we employ the standard Green's function technique to determine the first-order self-energy and from which we obtain renormalized energy spectrums. The self-energy we consider here is of the Hartree type and the corresponding Feynman diagram is shown in Fig.~\ref{fig:1}(c)~\cite{PhysRevX.8.041028}. The self-energy can be expressed as,
\begin{equation}
    \label{self-energy}
    \Sigma_{\alpha\gamma}(\mathbf{k},T)= \frac{1}{4SN\hslash} \sum\limits_{\mathbf{q},\lambda}S^{\lambda}_{\alpha\gamma}(\mathbf{k}, \mathbf{q})n_{\lambda}\left(\mathbf{q},T\right),
\end{equation}
where 
$\alpha$ and $\gamma$ represent either the up ($u$) or down ($d$) mode and $S^{\lambda}_{\alpha\gamma}(\mathbf{k}, \mathbf{q})$ is a function of momenta $\mathbf{k}$ and $\mathbf{q}$. It is understood from the Feynman diagram that the subscripts and the momentum of the self-energy correspond to those of the external Green's functions. The derivation of Eq.~(\ref{self-energy}) is in the supplementary material.


\par Our next step follows from the Dyson's equation,  $\mathcal{G}^{-1}\left({\mathbf{k}},\omega\right)=i\omega\mathds{1}-\hslash^{-1}\boldsymbol{\varepsilon}\left(\mathbf{k}\right)-\boldsymbol{\Sigma}\left({\mathbf{k}},\omega\right)$, where $\mathcal{G}^{-1}$ is the inverse of the dressed Green's function, $\omega$ is the bosonic Matsubara frequency,
$\boldsymbol{\varepsilon}\left({\mathbf{k}}\right)$ is the diagonalized matrix of $H_0$,
and $\boldsymbol{\Sigma}\left({\mathbf{k}},\omega\right)$ is the self-energy matrix.
It suggests that the effect of $\mathcal{H}_{\rm int}$ can be integrated into an effective single-particle Hamiltonian by adding the contribution from the self-energy, given in Eq. (\ref{self-energy}), to $\mathcal{H}_{0}$. That is,
\begin{equation}
    \label{Hud}
{\cal H}_{{\rm eff}}=\sum_{{\bf k}}\left(\begin{array}{cc}
\hat{u}_{{\bf k}}^{\dagger} & \hat{d}_{{\bf k}}^{\dagger}\end{array}\right)\left(\begin{array}{cc}
\varepsilon_u+\hbar\Sigma_{uu} & \hbar\Sigma_{ud}\\
\hbar\Sigma_{du} & \varepsilon_d+\hbar\Sigma_{dd}
\end{array}\right)\left(\begin{array}{c}
\hat{u}_{{\bf k}}\\
\hat{d}_{{\bf k}}
\end{array}\right).
\end{equation}
By directly diagonalizing the above effective Hamiltonian, we obtain the renormalized energy bands arising from the many-body correlation. We note  here that before evaluating the thermal Hall conductivity of our setup, one needs to rewrite $\cal{H}_{\rm eff}$ with the original $ab$-basis, that is, 
${\cal H}_{\rm{eff}}=\sum_{\mathbf{k}}\boldsymbol{\Psi^{\dagger}_{\mathbf{k}}}H_{\rm{eff}}(\mathbf{k})\boldsymbol{\Psi_{\mathbf{k}}}$.
\begin{figure*}[htbp] 
\includegraphics[width=7in,clip]{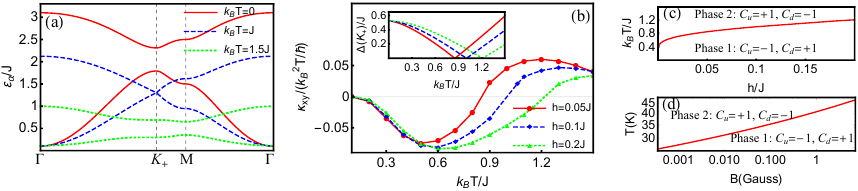}
\caption{
(a) Magnonic bandstructures for three different temperatures along the path $\Gamma-K_+-M-\Gamma$. Here $K_+$ is a Dirac point. 
(b) The thermal Hall conductivities vs temperature for three different Zeeman fields  are plotted. The connecting lines are guides to the eye.  The inset shows the gap $\Delta$ at the Dirac point $K_+$ as a function of temperature for the same three Zeeman fields. (c) The phase diagram as a function of temperature and Zeeman field for the same parameters as in (a) and (b). (d) The log-log plot of the phase diagram for $\rm{CrI_3}$ as a function of temperature and Zeeman field (see main text). 
}
\label{fig:2}
\end{figure*}
There are some intriguing phenomena occurred as a consequence of incorporating the first-order self energy. One of the major results in the paper is the recovery of the sign reversal of the thermal Hall conductivity when the magnon-magnon interactions are included. Below we shall present and discuss our theoretical results in detail. 

\textit{
Results and discussion.$-$}In this work, the following parameters are fixed unless otherwise stated: $S=\frac{1}{2}$ and $J^\prime=D=0.1J$~\cite{Supplement}. Since  Curie temperature ($T_{\rm{Curie}}$) is normally a few times larger than $k_B^{-1}J$~\cite{Onose297,PhysRevX.8.041028,Huang2017,PhysRevB.104.L020402}, the temperature  considered ranges from $k_BT=0$ to $k_BT=2J$. 
In Fig.~\ref{fig:2}(a), we plot  dispersion relations at three different temperatures when $h=0.1J$. As can be seen from Eq.~(\ref{self-energy}), the self-energy effect vanishes at $T=0$ because the Bose function for nonzero energies at low temperature approaches zero. As a result, the renormalized energies at $T=0$ are the same as the bare energies of Eq.~(\ref{single-particle_magnon_energies}) reported in Ref.~\cite{Owerre_2016}. As  the temperature increases, the gap at the Dirac point $K_+$ decreases and the spectrum becomes gapless at $T\approx J$ for this particular $h$. If we further increase the temperature, the gap reopens and its size increases with $T$. Note here that the magnonic excitation energy at $\Gamma$ is given by $h$ and is independent of the temperature. This is because the Zeeman interaction in Eq.~(\ref{Hamiltonian}) only couples to the $\hat{S}_z$ term which is exact in the HP transformation. 


To see the temperature dependence of the gap at $K_+$, we extract the gap for three different Zeeman energies as shown in the inset of Fig.~\ref{fig:2}(b). We find that as $h$ increases, the gap-closing temperature, $T_c$, also increases. Although only the gap at $K_+$ is shown, it is in fact the same as the gap at $K_-$ because of the inversion symmetry. It is straightforward to  show explicitly that both the perturbed and unperturbed Hamiltonian
possess the inversion symmetry by examining the relation $\sigma_xH\left(-\bf{k}\right)\sigma_x=H\left(\bf{k}\right)$ where $H=H_0$ or $H=H_{\rm eff}$. 
One can also draw the same conclusion from the fact that the magnetic field we consider here is out-of-plane and does not break the inversion symmetry. On the other hand, the orbital time-reversal symmetry condition $H^\ast\left(-\bf{k}\right)=H\left(\bf{k}\right)$~\cite{PhysRevX.8.011010} does not hold because of the presence of the DMI. 
Consequently, the gap sizes at $K_+$ and $K_-$ are identical but the gapless feature is no longer protected. 

The gap reopens at $T=T_c$ suggests that there may be a topological phase transition at $T_c$. To confirm this, we compute the Chern number of each band above and below $T_c$  and verify that it indeed changes sign at $T_c$. For definiteness, we plot in  Fig.~\ref{fig:2}(c) the phase boundary between these two distinct topological phases where the Chern numbers 
in phase 1 (phase 2)  are $C_u=-1\,(+1)$ and $C_d=+1\,(-1)$. The phase boundary in the $h$-$T$ phase diagram is determined by studying the gap at $K_+$ as in the inset of  Fig.~\ref{fig:2}(b). We note here that the Berry curvatures at $K_+$ and $K_-$ change sign simultaneously due to the inversion symmetry. This is in contrast to the Haldane model~\cite{PhysRevLett.61.2015} with a nonzero onsite energy.


As mentioned earlier, the topological phase transition can be experimentally confirmed from the sign reversal of the thermal Hall conductivity $\kappa_{xy}$. Owing to the two-band property that  $\Omega_u({\bf k})=-\Omega_d({\bf k})$, one can directly associate the thermal Hall conductivity in Eq.~(\ref{Hallconductivity}) to $\Omega_u\left(\bf{k}\right)$. In addition, we have checked numerically that $\Omega_u\left(\bf{k}\right)$ has the same sign as the Chern number $C_u$ for all $\bf{k}$ in the BZ.
From the fact that $c_2\left(x\right)$ is monotonic, one can deduce that $\kappa_{xy}$ changes sign when the magnonic system of honeycomb ferromagnets undergoes a topological phase transition. 
In Fig.~\ref{fig:2}(b), we plot $\kappa_{xy}$ as functions of $T$ for the same three Zeeman interactions in the inset. One can easily see that  $\kappa_{xy}$ evolves continuously with $T$ and goes from being negative at low $T$ ($T<T_c$) to being positive at high $T$ ($T>T_c$). 
Here, we provide evidences that the sign reversal of the transport quantity is clearly a strong indication for a topological phase transition. This also makes the honeycomb ferromagnet a good candidate for observing topological phase transitions in a bosonic system. We emphasize here that the sign reversal of $\kappa_{xy}$ does not share the same origin as that reported in Ref.~\cite{PhysRevLett.106.197202} where the gap does not close and reopen when the magnetic field continuously decreases to zero and changes its direction.

Qualitatively speaking, the new phenomenon we predict and the underlying physics
are rather generic and can be applied to real materials that are well described by the model Hamiltonian, Eq.~(\ref{Hamiltonian}), such as $\rm{CrBr_3}$ and $\rm{CrI_3}$~\cite{PhysRevX.8.011010,PhysRevX.8.041028,PhysRevB.104.L020402}. To  connect our theory with experiments, 
we take $\rm{CrI_3}$ as an example and adopt the following parameters obtained from fitting to experiments~\cite{PhysRevX.8.041028}: $S=3/2$, $J=2.01\rm{meV}$, $J^\prime\approx0.08J$, $D\approx0.15J$~\cite{jpp}. We find that, as shown in Fig.~\ref{fig:2}(d), below $T_{\rm{Curie}}\approx 45\rm{K}$ topological phase transitions occur with a critical magnetic field small compared to the case with $S=\frac{1}{2}$.
Specifically, the critical magnetic field for $h=0.1J$ in Fig.~\ref{fig:2}(c) is in the order of 1T if the same strength of $J$ is assumed there. Nevertheless, our theory suggests that the sign flip phenomenon of the thermal Hall conductivity  for $\rm{CrI_3}$ is experimentally accessible in the low $B$ limit by varying the temperature~\cite{PhysRevLett.115.106603}.


From the topological point of view, the DMI plays an essential role as opening up a gap for the magnonic bands. These non-crossings are the origin of a nonzero and well-defined Berry curvature and lead to nonzero topological invariants, i.e., Chern numbers. It is similar to the role of a spin-orbit interaction in  electronic topological insulators~\cite{RevModPhys.82.3045}. The major difference is that the topological phases of our system are characterized by the sign of the thermal Hall conductivity in contrast to the quantized Hall conductivity in electronic quantum Hall systems. It is worth mentioning that magnonic integer quantum Hall effects are also predicted to exist in a 2D spin-ice model under an out-of-plane field~\cite{PhysRevB.94.220403} or in a 2D clean insulating magnet under a skewed-harmonic electric potential where the quantized Hall conductance is associated with the Chern number for almost flat bands~\cite{PhysRevB.95.125429}.


Recently, the topological properties of two-dimensional square-lattice ferromagnets derived from a magnon-phonon coupling have also been investigated~\cite{PhysRevLett.123.237207,PhysRevLett.124.147204,PhysRevB.104.045139}. In particular, the phonons play a role when addressing the $\Gamma$-point physics~\cite{PhysRevLett.123.237207}. Since the topological properties in honeycomb ferromagnets are mainly determined by the Dirac-point physics, we do not expect our results will be drastically changed when the magnon-phonon coupling is considered. In addition, it is argued in Ref.~\cite{PhysRevB.104.045139} that an in-plane DMI in honeycomb magnets, which is absent in our setup, is directly responsible for the coupling between magnons and phonons. Nevertheless, incorporating a magnon-phonon coupling in our formalism may open a new venue in the field.


\textit{Conclusion.$-$}To summarize, 
when incorporating the Hartree-type self-energy related to the magnon-magnon interactions into the single-particle Hamiltonian,
the system possesses two topological phases, $(C_{u}=-1, C_{d}=+1)$ and $(C_{u}=+1, C_{d}=-1)$. The topological phase of the system can be tuned either by the temperature or external magnetic field. The corresponding continuous topological phase transitions driven by these magnon-magnon interactions are accompanied with a gap-closing phenomenon. The sign of the thermal Hall conductivity directly reflects the Chern numbers that characterize the topological phase. Therefore, the reversal of the sign  occurs during topological phase transitions and can serve as an indicator in future experiments. We note here that similar results are also reported in Ref.~\cite{PhysRevX.11.021061} where particle-number-nonconserving interactions are the key ingredients.
\acknowledgements{
This work is supported by MOST Grant No. 108-2112-M-009-004-MY3. We thank Chien-Hung Lin for illuminating conversations.
}
\bibliography{Review}

\begin{thebibliography}{40}
\bibitem [{jpp()}]{jpp}%
  \BibitemOpen
  \href@noop {} {}\bibinfo {note} {Since we expect the
  next-next-nearest-neighbor exchange interaction does not appreciably change
  our numerical results, it is not included here for simplicity.}\BibitemShut
  {Stop}%
\end{thebibliography}
\clearpage


\section*{Supplementary material}

# Supplementary Material: Topological Phase transitions of Dirac Magnons in Honeycomb Ferromagnets

Yu-Shan Lu,[1] Jian-Lin Li,[1] and Chien-Te Wu[1, 2]

[1]*Department of Electrophysics, National Yang Ming Chiao Tung University, Hsinchu, Taiwan*

[2]*Physics Division, National Center for Theoretical Sciences, Taipei, Taiwan*

## A. COEFFICIENTS OF MAGNON-MAGNON INTERACTION

In deriving the first-order self-energy, we take advantage of the orthonormal relation in the $ud$ basis to considerably simplify the process. In this section, we first derive the coefficients of the two-particle interacting Hamiltonian, $\mathcal{H}_{\text{int}}$, in Eq. (5) of the main text. Because there are 112 terms in total to deal with, the procedure is quite lengthy even though the principle behind the derivation is quite simple.

Using the Holstein-Primakoff (HP) transformation to expand the original Hamiltonian to the order of $S^{-\frac{1}{2}}$, we obtain the two-particle interacting Hamiltonian in the original $ab$ basis

$$\begin{aligned}
\mathcal{H}_{\text{int}} = & \frac{v_s}{4SN} \sum_{\{\mathbf{k}_i\}} \left( \gamma_{\mathbf{k}_4-\mathbf{k}_1-\mathbf{k}_2} \hat{b}^\dagger_{\mathbf{k}_1} \hat{b}^\dagger_{\mathbf{k}_2} \hat{a}_{\mathbf{k}_3} \hat{b}_{\mathbf{k}_4} + \gamma_{-\mathbf{k}_2} \hat{a}^\dagger_{\mathbf{k}_1} \hat{b}^\dagger_{\mathbf{k}_2} \hat{a}_{\mathbf{k}_3} \hat{a}_{\mathbf{k}_4} + \gamma_{\mathbf{k}_3} \hat{a}^\dagger_{\mathbf{k}_1} \hat{a}^\dagger_{\mathbf{k}_2} \hat{b}_{\mathbf{k}_3} \hat{a}_{\mathbf{k}_4} + \gamma_{-\mathbf{k}_1+\mathbf{k}_4+\mathbf{k}_3} \hat{b}^\dagger_{\mathbf{k}_1} \hat{a}^\dagger_{\mathbf{k}_2} \hat{b}_{\mathbf{k}_3} \hat{b}_{\mathbf{k}_4} \right. \\
& \left. -4\gamma_{-\mathbf{k}_2+\mathbf{k}_4} \hat{a}^\dagger_{\mathbf{k}_1} \hat{b}^\dagger_{\mathbf{k}_2} \hat{a}_{\mathbf{k}_3} \hat{b}_{\mathbf{k}_4} \right) \delta_{-\mathbf{k}_1-\mathbf{k}_2+\mathbf{k}_3+\mathbf{k}_4,0} \\
& + \frac{1}{4SN} \sum_{\{\mathbf{k}_i\}} \left\{ \left[ v_t \left( \cos\phi \, \mathcal{P}_{\{\mathbf{k}_i\}} - \sin\phi \, \varrho_{\{\mathbf{k}_i\}} \right) - 4v'_s \, p_{\mathbf{k}_4-\mathbf{k}_2} \right] \hat{a}^\dagger_{\mathbf{k}_1} \hat{a}^\dagger_{\mathbf{k}_2} \hat{a}_{\mathbf{k}_3} \hat{a}_{\mathbf{k}_4} \right. \\
& \left. + \left[ v_t \left( \cos\phi \, \mathcal{P}_{\{\mathbf{k}_i\}} + \sin\phi \, \varrho_{\{\mathbf{k}_i\}} \right) - 4v'_s \, p_{\mathbf{k}_4-\mathbf{k}_2} \right] \hat{b}^\dagger_{\mathbf{k}_1} \hat{b}^\dagger_{\mathbf{k}_2} \hat{b}_{\mathbf{k}_3} \hat{b}_{\mathbf{k}_4} \right\} \delta_{-\mathbf{k}_1-\mathbf{k}_2+\mathbf{k}_3+\mathbf{k}_4,0},
\end{aligned} \quad (\text{S1})$$

where $\mathcal{P}_{\{\mathbf{k}_i\}} \equiv p_{\mathbf{k}_1+\mathbf{k}_2-\mathbf{k}_4} + p_{\mathbf{k}_4+\mathbf{k}_3-\mathbf{k}_1} + p_{\mathbf{k}_2} + p_{\mathbf{k}_3}$ and $\varrho_{\{\mathbf{k}_i\}} \equiv \rho_{\mathbf{k}_1+\mathbf{k}_2-\mathbf{k}_4} + \rho_{\mathbf{k}_4+\mathbf{k}_3-\mathbf{k}_1} + \rho_{\mathbf{k}_2} + \rho_{\mathbf{k}_3}$. In order to use the orthonormal relations in the $ud$ basis, we transform the two-particle interaction from the $ab$ basis to the $ud$ basis with the following relations,

$$\hat{a}_{\mathbf{k}} = \frac{1}{\sqrt{2}} \left( A(\mathbf{k}) \hat{u}_{\mathbf{k}} + M(\mathbf{k}) \hat{d}_{\mathbf{k}} \right), \quad (\text{S2a})$$

$$\hat{b}_{\mathbf{k}} = \frac{e^{-i\Phi(\mathbf{k})}}{\sqrt{2}} \left( M(\mathbf{k}) \hat{u}_{\mathbf{k}} - A(\mathbf{k}) \hat{d}_{\mathbf{k}} \right), \quad (\text{S2b})$$

where $A(\mathbf{k}) \equiv \sqrt{1 + \frac{h_z(\mathbf{k})}{\epsilon(\mathbf{k})}}$ and $M(\mathbf{k}) \equiv \sqrt{1 - \frac{h_z(\mathbf{k})}{\epsilon(\mathbf{k})}}$. Substituting Eqs. (S2a) and (S2b) into Eq. (S1), one can derive the explicit forms of $V_{1,\{\mathbf{k}_i\}}$ to $V_{9,\{\mathbf{k}_i\}}$ defined in the main text.

## B. COEFFICIENTS OF THE FIRST-ORDER SELF-ENERGY

The first-order self-energy can be expressed as $\Sigma^{(1)}_{\alpha\gamma}(\mathbf{k}, T) = \frac{1}{4SN\hbar} \sum_{\mathbf{q}} \left( S^u_{\alpha\gamma}(\mathbf{k}, \mathbf{q}) n_u(\mathbf{q}, T) + S^d_{\alpha\gamma}(\mathbf{k}, \mathbf{q}) n_d(\mathbf{q}, T) \right)$. Due to the orthonormal relation, it can be shown that $V_{3,\{\mathbf{k}\}}$ and $V_{7,\{\mathbf{k}\}}$ do not contribute to the first-order Green's function. Moreover, the three contractions yields three Kronecker deltas, which get rid of three momentum sums.

The following are the forms of $S^\lambda_{\alpha\gamma}(\mathbf{k},\mathbf{q})$:

$$\begin{cases} S^u_{uu}(\mathbf{k},\mathbf{q}) = & \frac{1}{4SN\hbar}[V_1(\mathbf{k}_1=\mathbf{k}_4=\mathbf{k};\mathbf{k}_2=\mathbf{k}_3=\mathbf{q}) + V_1(\mathbf{k}_1=\mathbf{k}_3=\mathbf{k};\mathbf{k}_2=\mathbf{k}_4=\mathbf{q}) \\ & + V_1(\mathbf{k}_2=\mathbf{k}_4=\mathbf{k};\mathbf{k}_1=\mathbf{k}_3=\mathbf{q}) + V_1(\mathbf{k}_2=\mathbf{k}_3=\mathbf{k};\mathbf{k}_1=\mathbf{k}_4=\mathbf{q})] \\ S^d_{uu}(\mathbf{k},\mathbf{q}) = & \frac{1}{4SN\hbar}[V_5(\mathbf{k}_1=\mathbf{k}_3=\mathbf{k};\mathbf{k}_2=\mathbf{k}_4=\mathbf{q})] \\ S^u_{ud}(\mathbf{k},\mathbf{q}) = & \frac{1}{4SN\hbar}[V_2(\mathbf{k}_2=\mathbf{k}_4=\mathbf{k};\mathbf{k}_1=\mathbf{k}_3=\mathbf{q}) + V_2(\mathbf{k}_1=\mathbf{k}_4=\mathbf{k};\mathbf{k}_2=\mathbf{k}_3=\mathbf{q}) \\ S^d_{ud}(\mathbf{k},\mathbf{q}) = & \frac{1}{4SN\hbar}[V_6(\mathbf{k}_1=\mathbf{k}_4=\mathbf{k};\mathbf{k}_2=\mathbf{k}_3=\mathbf{q}) + V_6(\mathbf{k}_1=\mathbf{k}_3=\mathbf{k};\mathbf{k}_2=\mathbf{k}_4=\mathbf{q}) \\ S^u_{du}(\mathbf{k},\mathbf{q}) = & \frac{1}{4SN\hbar}[V_4(\mathbf{k}_2=\mathbf{k}_4=\mathbf{k};\mathbf{k}_1=\mathbf{k}_3=\mathbf{q}) + V_4(\mathbf{k}_2=\mathbf{k}_3=\mathbf{k};\mathbf{k}_1=\mathbf{k}_4=\mathbf{q})] \\ S^d_{du}(\mathbf{k},\mathbf{q}) = & \frac{1}{4SN\hbar}[V_8(\mathbf{k}_2=\mathbf{k}_3=\mathbf{k};\mathbf{k}_1=\mathbf{k}_4=\mathbf{q}) + V_8(\mathbf{k}_1=\mathbf{k}_3=\mathbf{k};\mathbf{k}_2=\mathbf{k}_4=\mathbf{q})] \\ S^u_{dd}(\mathbf{k},\mathbf{q}) = & \frac{1}{4SN\hbar}[V_5(\mathbf{k}_2=\mathbf{k}_4=\mathbf{k};\mathbf{k}_1=\mathbf{k}_3=\mathbf{q})] \\ S^d_{dd}(\mathbf{k},\mathbf{q}) = & \frac{1}{4SN\hbar}[V_9(\mathbf{k}_2=\mathbf{k}_4=\mathbf{k};\mathbf{k}_1=\mathbf{k}_3=\mathbf{q}) + V_9(\mathbf{k}_2=\mathbf{k}_3=\mathbf{k};\mathbf{k}_1=\mathbf{k}_4=\mathbf{q}) \\ & + V_9(\mathbf{k}_1=\mathbf{k}_4=\mathbf{k};\mathbf{k}_2=\mathbf{k}_3=\mathbf{q}) + V_9(\mathbf{k}_1=\mathbf{k}_3=\mathbf{k};\mathbf{k}_2=\mathbf{k}_4=\mathbf{q})] \end{cases}. \quad (S3)$$

The explicit expressions of $S^u_{\alpha\gamma}$ and $S^d_{\alpha\gamma}$ in Eqs. (S3) are

$$\begin{aligned} 4SN\hbar S^{u,d}_{uu}(\mathbf{k},T) =& 2\left[-\frac{h_x(\mathbf{k})^2+h_y(\mathbf{k})^2}{\epsilon(\mathbf{k})} \mp \frac{h_x(\mathbf{q})^2+h_y(\mathbf{q})^2}{\epsilon(\mathbf{q})}\right] \\ &+ 2v_s\left[\mp\text{Re}\{e^{i(\Phi_\mathbf{q}-\Phi_\mathbf{k})}\gamma_{\mathbf{k}-\mathbf{q}}\}\sqrt{1-\frac{h_z(\mathbf{k})^2}{\epsilon(\mathbf{k})^2}}\sqrt{1-\frac{h_z(\mathbf{q})^2}{\epsilon(\mathbf{q})^2}} - 3\left(1\mp\frac{h_z(\mathbf{k})h_z(\mathbf{q})}{\epsilon(\mathbf{k})\epsilon(\mathbf{q})}\right)\right] \\ &+ 4\left[v_t\cos\phi(p_\mathbf{q}+p_\mathbf{k}) - v'_s p_{\mathbf{k}-\mathbf{q}} - 3v'_s\right]\left(1\pm\frac{h_z(\mathbf{k})h_z(\mathbf{q})}{\epsilon(\mathbf{k})\epsilon(\mathbf{q})}\right) \\ &+ 4v_t\sin\phi(\rho_\mathbf{q}+\rho_\mathbf{k})\left(-\frac{h_z(\mathbf{k})}{\epsilon(\mathbf{k})} \mp \frac{h_z(\mathbf{q})}{\epsilon(\mathbf{q})}\right), \end{aligned} \quad (S4)$$

$$\begin{aligned} 4SN\hbar S^{u,d}_{ud} =& 2\frac{\sqrt{h_x(\mathbf{k})^2+h_y(\mathbf{k})^2}h_z(\mathbf{k})}{\epsilon(\mathbf{k})} \pm 6v_s\sqrt{1-\frac{h_z(\mathbf{k})^2}{\epsilon(\mathbf{k})^2}}\frac{h_z(\mathbf{q})}{\epsilon(\mathbf{q})} \\ &+ 2v_s\left[\pm\frac{h_z(\mathbf{k})}{\epsilon(\mathbf{k})}\text{Re}\{e^{i(\Phi_\mathbf{k}-\Phi_\mathbf{q})}\gamma_{-\mathbf{k}+\mathbf{q}}\} \mp i\text{Im}\{e^{i(\Phi_\mathbf{k}-\Phi_\mathbf{q})}\gamma_{-\mathbf{k}+\mathbf{q}}\}\right]\sqrt{1-\frac{h_z(\mathbf{q})^2}{\epsilon(\mathbf{q})^2}} \\ &+ 4\sqrt{1-\frac{h_z(\mathbf{k})^2}{\epsilon(\mathbf{k})^2}}\left\{-v_t\sin\phi(\rho_k+\rho_q) \pm \frac{h_z(\mathbf{q})}{\epsilon(\mathbf{q})}\left[v_t\cos\phi(p_\mathbf{k}+p_\mathbf{q}) - v'_s p_{\mathbf{k}-\mathbf{q}} - 3v'_s\right]\right\}, \end{aligned} \quad (S5)$$

$$S^{u,d}_{du} = (S^{u,d}_{ud})^*, \quad (S6)$$

$$\begin{aligned} 4SN\hbar S^{u,d}_{dd} =& 2\left[\frac{h_x(\mathbf{k})^2+h_y(\mathbf{k})^2}{\epsilon(\mathbf{k})} \mp \frac{h_x(\mathbf{q})^2+h_y(\mathbf{q})^2}{\epsilon(\mathbf{q})}\right] + 2v_s\left[\pm\text{Re}\{e^{i(\Phi_\mathbf{q}-\Phi_\mathbf{k})}\gamma_{\mathbf{k}-\mathbf{q}}\}\sqrt{1-\frac{h_z(\mathbf{k})^2}{\epsilon(\mathbf{k})^2}}\sqrt{1-\frac{h_z(\mathbf{q})^2}{\epsilon(\mathbf{q})^2}}\right. \\ &\left.- 3\left(1\pm\frac{h_z(\mathbf{k})h_z(\mathbf{q})}{\epsilon(\mathbf{k})\epsilon(\mathbf{q})}\right)\right] + 4\left[v_t\cos\phi(p_\mathbf{q}+p_\mathbf{k}) - v'_s p_{\mathbf{k}-\mathbf{q}} - 3v'_s\right]\left(1\mp\frac{h_z(\mathbf{k})h_z(\mathbf{q})}{\epsilon(\mathbf{k})\epsilon(\mathbf{q})}\right) \\ &+ 4v_t\sin\phi(\rho_q+\rho_k)\left[\frac{h_z(\mathbf{k})}{\epsilon(\mathbf{k})} \mp \frac{h_z(\mathbf{q})}{\epsilon(\mathbf{q})}\right], \end{aligned} \quad (S7)$$

where the upper and lower signs are for $S^u_{\alpha\gamma}$ and $S^d_{\alpha\gamma}$, respectively.

### C. AVOID UNDEFINED PHASE IN THE EIGENSTATES FOR DIRAC POINTS

The phase in Eq. (S2b) are not defined at two Dirac points in Brillouin zone (BZ) owing to the fact that $h_x(\mathbf{K}_\pm) = h_y(\mathbf{K}_\pm) = 0$. Therefore, one can not compute relevant physical quantities such as the self-energy and effective

Hamiltonian with this particular gauge. To resolve this problem, we divide the BZ into two regions, each one contains a single Dirac point. In the region containing $\mathbf{K}_+$, we use

$$\hat{a}_\mathbf{k} = \frac{1}{\sqrt{2}} \left( \frac{h_x(\mathbf{k}) + ih_y(\mathbf{k})}{\sqrt{\epsilon(\mathbf{k})(\epsilon(\mathbf{k}) - h_z(\mathbf{k}))}} \hat{u}_\mathbf{k} + \sqrt{1 - \frac{h_z(\mathbf{k})}{\epsilon(\mathbf{k})}} \hat{d}_\mathbf{k} \right),$$
$$\hat{b}_\mathbf{k} = \frac{1}{\sqrt{2}} \left( \sqrt{1 - \frac{h_z(\mathbf{k})}{\epsilon(\mathbf{k})}} \hat{u}_\mathbf{k} - \frac{h_x(\mathbf{k}) - ih_y(\mathbf{k})}{\sqrt{\epsilon(\mathbf{k})(\epsilon(\mathbf{k}) - h_z(\mathbf{k}))}} \hat{d}_\mathbf{k} \right), \quad \text{(S8)}$$

corresponding to a gauge different than Eq. (S2b). While in the other region containing $\mathbf{K}_-$, we adopt another gauge to write down

$$\hat{a}_\mathbf{k} = \frac{1}{\sqrt{2}} \left( \sqrt{1 + \frac{h_z(\mathbf{k})}{\epsilon(\mathbf{k})}} \hat{u}_\mathbf{k} - \frac{h_x(\mathbf{k}) + ih_y(\mathbf{k})}{\sqrt{\epsilon(\mathbf{k})(\epsilon(\mathbf{k}) + h_z(\mathbf{k}))}} \hat{d}_\mathbf{k} \right),$$
$$\hat{b}_\mathbf{k} = \frac{1}{\sqrt{2}} \left( \frac{h_x(\mathbf{k}) - ih_y(\mathbf{k})}{\sqrt{\epsilon(\mathbf{k})(\epsilon(\mathbf{k}) + h_z(\mathbf{k}))}} \hat{u}_\mathbf{k} + \sqrt{1 + \frac{h_z(\mathbf{k})}{\epsilon(\mathbf{k})}} \hat{d}_\mathbf{k} \right). \quad \text{(S9)}$$

Both Eqs. (S8) and (S9) are used to perform momentum sums when evaluating physical quantities of interest. Although we use different gauges to compute the physical quantities, one can prove that the self-energy in the $ab$ basis, effective Hamiltonian, energy bands, Berry curvature, and Hall conductivity are all gauge invariant.

## D. EFFECTS FROM THE DZYALOSHINSKII-MORIYA INTERACTION AND THE NEXT-NEAREST-NEIGHBOR EXCHANGE INTERACTION

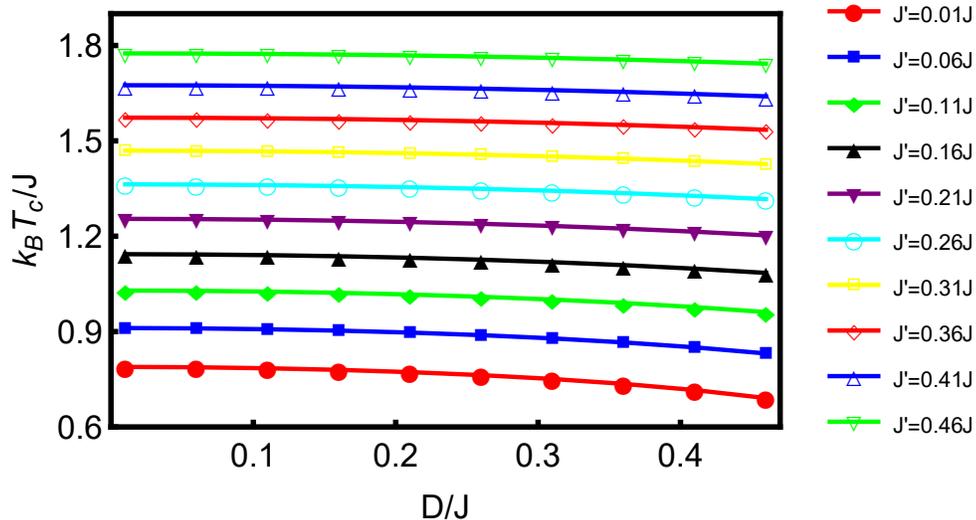

FIG. S1. The critical temperatures as a function of the Dzyaloshinskii-Moriya interaction, $D$, for different strengths of the next-nearest-neighbor exchange interaction, $J'$. The lines connecting symbols are guides for the eye.

We present numerical results on the effects of the Dzyaloshinskii-Moriya interaction (DMI) and the next-nearest-neighbor (NNN) exchange interaction, $J'$, in this section. The Zeeman interaction is fixed to be $0.1J$ and the magnitude of spin is $S = 1/2$ as in Figs. 2(a), 2(b), and 2(c) in the main text. In Fig. S1, the topological phase transition temperatures, $T_c$, are plotted against the DMI for a number of the NNN exchange interaction ranging from $J' = 0.01J$ to $J' = 0.46J$.

Before discussing details of this figure, it is important to note that these topological phase transitions are driven by the fluctuation of spin texture originating from magnon-magnon interactions. One anticipates that the DMI enhances the fluctuation while $J'$ suppresses it simply because $J'$ is positive and helps to stabilize the ferromagnetic phase.

With this in mind, one first see from Fig. S1 that the dependence of $T_c$ on the DMI is getting weaker when $J'$ is getting larger. For example, the curve for $J' = 0.01J$ drops more noticeably than the curve for $J' = 0.46J$. This can be understood from the role of $J'$. As it increases, the spin-texture is expected to become more uniform and the effect from the DMI is reduced.

Next, for a fixed ratio of $D/J$, $T_c$ increases steadily with $J'$. On the other hand, when $J'/J$ is fixed, $T_c$ does not drop significantly when the DMI is increased. Nevertheless, the effects of the NNN exchange interaction and DMI are only quantitative and topological phase transitions studied in the main text are a general phenomenon.

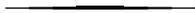


\clearpage

\end{document}